\documentclass[preprint,5p,times]{elsarticle}

\usepackage{amsfonts,amsmath,amsthm,amssymb}
\usepackage{graphicx,epsfig,pstricks,pst-node}
\usepackage{xcolor}

\journal{Chaos, Solitons and Fractals}

\begin{document}
	
\begin{frontmatter}
		
\title{Modulating biodiversity through higher-order interactions and intraspecific competition in rock-paper-scissors dynamics}

\author[ku]{Chunpeng Du}
\author[ku]{Haoshu Wang}
\author[yu,yusc]{Yikang Lu\corref{cor1}}
\ead{luyikang\_top@163.com}

\author[ku]{Lijuan Qin\corref{cor1}}
\ead{Km\_Qinlijuan@126.com}	
\author[khu,khu2]{Junpyo Park\corref{cor1}}
\ead{junpyopark@khu.ac.kr}

\cortext[cor1]{Corresponding authors}

\address[ku]{School of Mathematics, Kunming University, Kunming 650214, China}
\address[yu]{School of Statistics and Mathematics, Yunnan University of Finance and Economics, Kunming, Yunnan 650221, China}
\address[yusc]{Yunnan Key Laboratory of Service Computing, Yunnan University of Finance and Economics, Yunnan 650221, China}
\address[khu]{Department of Applied Mathematics, 
	Colleage of Applied Sciences,
	Kyung Hee University, 
	Yongin 17104, Republic of Korea}
\address[khu2]{G-LAMP NEXUS Institute, 
	Kyung Hee University, 
	Yongin 17104, Republic of Korea}

\date{\today}

\begin{abstract}
	Understanding the mechanisms that govern species coexistence and biodiversity represents a fundamental challenge in ecology. This study extends the classic rock-paper-scissors model by introducing a context-dependent higher-order interaction mechanism where intraspecific competition is dynamically regulated by local resource availability. Crucially, our quantitative analysis reveals that higher-order interactions significantly modulate the system's structural organization: Enhanced strength of higher-order interactions leads to a decrease in spatial wavelength, resulting in the formation of more compact species domains. However, this structural change makes the system more sensitive to mobility, shifting the extinction threshold to lower values. These findings highlight the dual role of resource-mediated higher-order regulation: it promotes local pattern formation but alters the system's resilience to dispersal, providing new theoretical perspectives for biodiversity conservation.
\end{abstract}


\end{frontmatter}

\section{\label{sec:intro}Introduction}

Coexistence and interaction among multispecies in ecosystems are at the core of biodiversity, and understanding the factors affecting biodiversity is thus a central issue in ecological sciences~\cite{1,2,3,4,5,6,7,8,9,10}.
To elucidate such complicated issues in reasonable and theoretical ways, evolutionary games have been a powerful framework and the rock-paper-scissors (RPS) game has served as a canonical model for investigating non-hierarchical cyclic interactions among species, such as microbial warfare among colicinogenic bacteria~\cite{11}, alternative mating strategies in side-blotched lizards~\cite{12}, competitive dynamics among yeast strains~\cite{13}, and community interactions within coral reef ecosystems~\cite{14}.

For the past few decades, uncovering the mechanisms that preserve biodiversity has been considered a key issue in ecological research. In particular, for systems governed by RPS games, researchers have increasingly examined multiple interacting factors that influence species coexistence and evolutionary patterns. 
A primary focus of most previous studies has been the dual role of individual mobility, as highlighted in a milestone work~\cite{40}. This role not only significantly influences species coexistence in spatially structured cyclic competition systems~\cite{15,16,17}, but also facilitates the emergence of cooperative behavior in networked environments~\cite{18}.
Beyond mobility, studies have investigated the impacts of intraspecific competition~\cite{20,21,22,23}, habitat suitability~\cite{24}, mutation dynamics~\cite{25,26}, higher-order interactions~\cite{34,35,36,37,38,39} and community size effects~\cite{19}.
Recent advances further reveal how cyclic dominance shapes broader ecological and evolutionary processes, from spatial pattern formation~\cite{27} to invasive dynamics in multispecies systems~\cite{28,29,30}, demonstrating the far-reaching implications of these interactions across ecological and social systems.

The traditional RPS model assumes that competition between species and within species is paired-interaction and homogeneous, with environmental resources, such as food, being supplied to the model in a constant manner~\cite{20,21}. It does not fully consider the deep regulation of competition mechanisms by variable resources and higher-order interactions. However, ecological factors, including food scarcity, spatial heterogeneity, and population density, can modulate competition intensity and strategic evolution through higher-order interactions~\cite{31}.
On the other hand, empirical studies in real systems such as side-blotched lizards~\cite{32} and microbial populations~\cite{11,13,14,33}, which are representative systems elucidated by the RPS game, demonstrate that fluctuating resource conditions can exacerbate intraspecific competition, driving adaptive behavioral shifts or strategic transitions.
To reflect and consider such deep regulatory properties, in recent years, higher-order interactions (HOI) have been demonstrated as a crucial mechanism for community stability and the formation of biodiversity in both theoretical and empirical ecology~\cite{34,35,36,37,38,39}. Unlike traditional pairwise models, HOI can disrupt cyclic dominance structures and induce complex ecological dynamics, including diversification, cooperation, and dissipation, under spatially explicit or resource-mediated conditions. Despite these insights, there remains a relative paucity of theoretical exploration regarding how 'food quantity' influences intraspecific competition within RPS systems through higher-order regulatory mechanisms, and how these mechanisms, in conjunction with individual mobility, regulate species coexistence. Mobility, a key ecological parameter that reflects dispersal capacity, shapes spatial structure and the intensity of population mixing. Its synergistic effect with higher-order regulatory factors may critically determine biodiversity patterns and extinction risks.

In this regard, we aim to investigate how higher-order interactions affect the maintenance of biodiversity and ecosystem stability through the mechanism of resource-mediated intraspecific competition regulation implemented at the individual level. Unlike previous studies focusing on general HOI frameworks, we specifically examine how local resource availability creates context-dependent regulatory effects on the competitive dynamics between neighboring individuals. Crucially, in our model, the outcome of a pairwise competitive encounter is co-determined by the collective state of the immediate local neighborhood surrounding the interacting pair, representing a microscopic, individual-based HOI. Through systematic numerical simulations of ecosystem evolutionary dynamics under varying levels of mobility and higher-order interaction strength, we identify fundamental findings. First, low and intermediate mobility regimes sustain robust species coexistence, characterized by stable spiral wave patterns. Second, high mobility consistently destabilizes the system, leading to biodiversity loss and eventual single-species dominance. Most importantly, higher-order interactions act as a crucial destabilizing modulator: increasing HOI strength significantly shifts the extinction threshold toward lower mobility.

The remainder of the paper is organized as follows. In Section~\ref{sec:model}, we introduce the spatial rock-paper-scissors game incorporating resource-mediated higher-order interactions for modeling intraspecific competition. In Sec. \ref{sec:results}, we present numerical results on population evolution, quantitative spatial pattern analysis, and extinction probability under this regulatory mechanism. Concluding remarks are addressed in Sec.~\ref{sec:conc}.

\section{\label{sec:model}Model Description}

\begin{figure}[ht]
	\centering
	\includegraphics[width=1\linewidth]{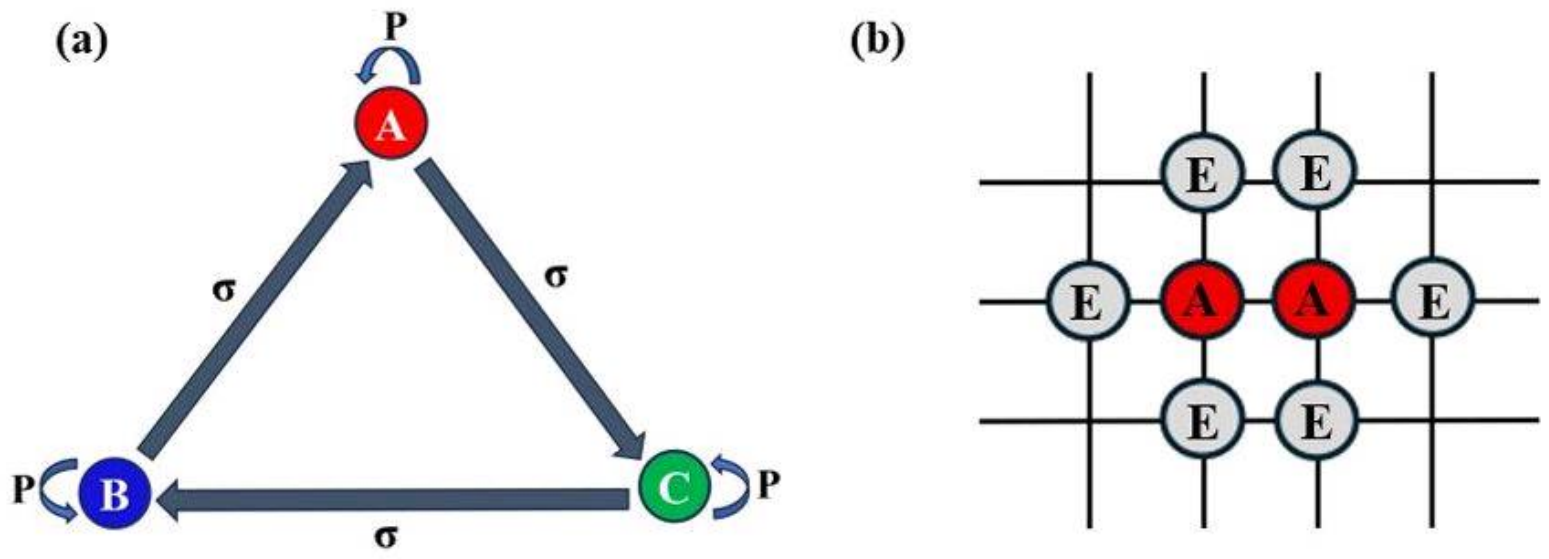}
	\caption{(a) Schematic diagram of the extended RPS model with intraspecific competition. Species $A$ (red), $B$ (blue), and $C$ (green) are engaged in cyclic dominance interactions. Based on the fundamental structure of a classic RPS system, each species can have intraspecific competition, which occurs depending on the food quantity and is determined by higher-level regulatory parameters dynamically. For details, refer to the model equation. (b) Geometric illustration of the local interaction neighborhood. The central adjacent pair represents the interacting individuals. $E$ represents any species or empty sites. The surrounding squares show the six unique nearest neighbors used to calculate the local resource abundance $\rho$, justifying the denominator in Eq.~(\ref{eqn:rate_P}).}
		\label{fig:1}
\end{figure}

We use the rock-paper-scissors game to model the cyclic interactions among three mobile species with intraspecific competition. 
In the spatial dynamics, three species $A$, $B$, and $C$ randomly populate sites on a square lattice of size $N = L \times L$ with periodic boundary conditions. The system evolves through asynchronous random sequential updates, and species interact with each other in the nearest neighborhood following the reaction rules:
\begin{eqnarray}
	&& AB \xrightarrow{\sigma} A \varnothing, \quad BC \xrightarrow{\sigma} B \varnothing, \quad CA \xrightarrow{\sigma} C \varnothing,  \label{eqn:1}\\
	&& A \varnothing \xrightarrow{\mu} AA, \quad B \varnothing \xrightarrow{\mu} BB, \quad C \varnothing \xrightarrow{\mu} CC,  \label{eqn:2}\\
	&& A \square \xrightarrow{r} \square A, \quad B \square \xrightarrow{r} \square B, \quad C \square \xrightarrow{r} \square C,  \label{eqn:3}\\
	&& AA \xrightarrow{P} A \varnothing, \quad BB \xrightarrow{P} B \varnothing, \quad CC \xrightarrow{P} C \varnothing,  \label{eqn:4}
\end{eqnarray}
where $A$, $B$, and $C$ denote three species engaged in cyclical competition, $\varnothing$ denotes an empty site by competition~(\ref{eqn:1}), and $\square$ describes any species or an empty site in the system. 
Reaction~(\ref{eqn:1})  represents interspecific competitions, which occur at the rate $\sigma$. For instance, if species $A$ competes with species $B$, the site occupied by $B$ transitions to an empty state. 
Reaction~(\ref{eqn:2}) denotes species reproduce by populating an adjacent empty site at a rate $\mu$. Reaction~(\ref{eqn:3}) describes the migration process by exchanging positions among nearest neighbors, which occurs at a rate of $r$ defined by $r=2MN$ with an individual’s mobility $M$ and size of a square lattice $N$ as usual~\cite{40,41}. Reaction~(\ref{eqn:4}) represents intraspecific competition that occurs between individuals of the same species. If individuals of the same species $A$ compete with each other, the winning individual will convert the node of the eliminated individual into an empty state. Unlike previous studies~\cite{21,23} that assumed constant rates, we introduce an intraspecific competition mechanism regulated by local resource availability in the immediate environment. The intraspecific competition rate is defined as follows:
\begin{equation}
	P = \alpha \exp\left(\dfrac{\rho}{6}\delta\right), \label{eqn:rate_P}
\end{equation}
where $\alpha$ represents the baseline intraspecific competition rate, $\rho$ denotes the local food resource abundance, defined as the count of food among the six nearest-neighbor positions surrounding the competing pair of individuals. The parameter $\delta$ serves as the higher-order modulation factor, controlling the sensitivity of competition to local resource levels. A schematic diagrams of the competition reactions (\ref{eqn:1}) and (\ref{eqn:4}) are presented in Fig.~\ref{fig:1}(a). The denominator $6$ in Eq.~(\ref{eqn:rate_P}) originates from the total number of lattice sites in the ``edge neighborhood'' centered on the competing pair. Specifically, on a square grid, the local neighborhood of a competing pair contains $3 + 3 = 6$ distinct nearest-neighbor sites. This mechanism implements a neighborhood mediated higher-order interaction: the competitive outcome between a pair of individuals is functionally co-determined by the collective state of their local neighborhood group, specifically through the integrated influence of all six surrounding neighboring sites as illustrated in Fig.~\ref{fig:1}(b).

To be concrete, at each simulation step, a randomly chosen individual interacts with one of its nearest neighbors at random. For the pair of selected nodes, intraspecific competition, interspecific competition, reproduction, and exchange occur with normalized probabilities: $P/(\sigma+\mu+P+r)$, $\sigma/(\sigma+\mu+P+r)$, $\mu/(\sigma+\mu+P+r)$, and $r/(\sigma+\mu+P+r)$.
After a random pair of nearest neighboring sites is selected, a chosen reaction is performed, if it is allowed. For example, if reproduction is chosen but there are no empty sites, the reaction fails.
We carry out simulations for a typical waiting time $T=32N$ until extinction occurs for the high-mobility regime and use the same time for all regimes of mobility considered in this paper. To make an unbiased comparison with previous works\cite{40}, we assume equal reaction probabilities for reproduction and interspecific competition, i.e., $\sigma=\mu=1$. Our analysis focuses on the effects introduced by the higher-order interaction term ($\delta$) in Eq.~(\ref{eqn:rate_P}). To this end, we first establish a baseline ($\delta=0$) that replicates key behaviors of the classic RPS system. As a set of free parameters in this system, to explore evolutionary dynamics of the system engaging with higher-order interaction on intraspecific competition, we set the baseline value $\alpha$ as $\alpha = 0.1$,which exhibits a more closely related behavior in the absence of higher-order intraspecific competition\cite{letten2019mechanistic}.

To ensure full reproducibility and methodological transparency, we provide a complete model description following the standard ODD (Overview, Design concepts, Details) protocol~\cite{Grimm2010} in Supplementary Material.

\section{\label{sec:results}Result}

The prevalence of higher-order interactions among species warrants thorough investigation. Therefore, we investigate the impact of food quantity on intraspecific competition under higher-order interactions from both microscopic phenomena and macroscopic perspectives in species evolutionary dynamics.

\subsection{\label{subsec:micro_sim}Microscopic phenomena in the evolutionary dynamics of species}

\begin{figure}[ht]
	\centering
	\includegraphics[width=0.9\linewidth]{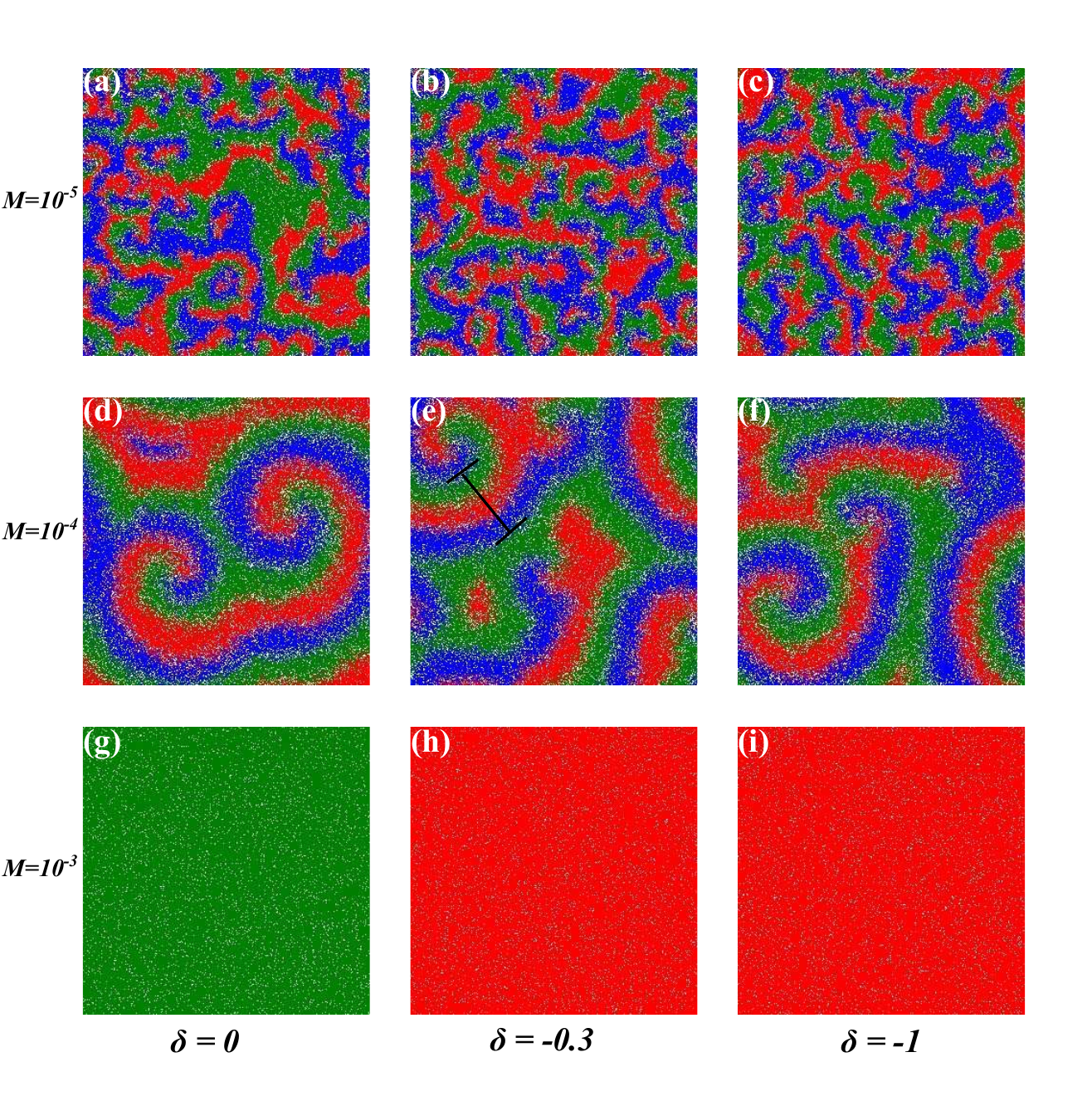}
	\caption{Characteristic snapshots with different combinations of $M$ and $\delta$ at the 180000 time step. 
		Panels from tops to bottoms are considered with different $M$: $M = 10^{-5}$, $10^{-4}$, and $10^{-3}$, respectively.
		In each row, panels from left to right are obtained with different $\delta$: $\delta=0$, $-0.3$, and $-1$, respectively.
		The colors red, blue, and green represent species $A$, $B$, and $C$, respectively. 
		These snapshots illustrate the typical spatial patterns arising after long-term evolution under different combinations of species mobility and the strength of high-order interactions. The overall survival patterns are similar to those in the classic model. A remarkable point is that, even if a single species dominates the system, the spatial system is not fully occupied by the species; instead, it survives with emerging empty sites.}
	\label{fig:2}
\end{figure}

We first explore the impact of higher-order interactions on species biodiversity at the microscopic level. To elucidate pattern formations and the presence of empty sites, we employ an $N = 300 \times 300$ sized lattice network, where three species and empty sites are initially randomly distributed across the network. 
To investigate the influence of mobility and the higher-order interaction parameter $\delta$, we first analyze characteristic snapshots of different steady states under varying $\delta$ values at fixed mobility values of $10^{-5}$, $10^{-4}$, and $10^{-3}$, which are presented in Fig.~\ref{fig:2}.

\begin{figure*}[ht]
	\centering
	\includegraphics[width=0.75\linewidth]{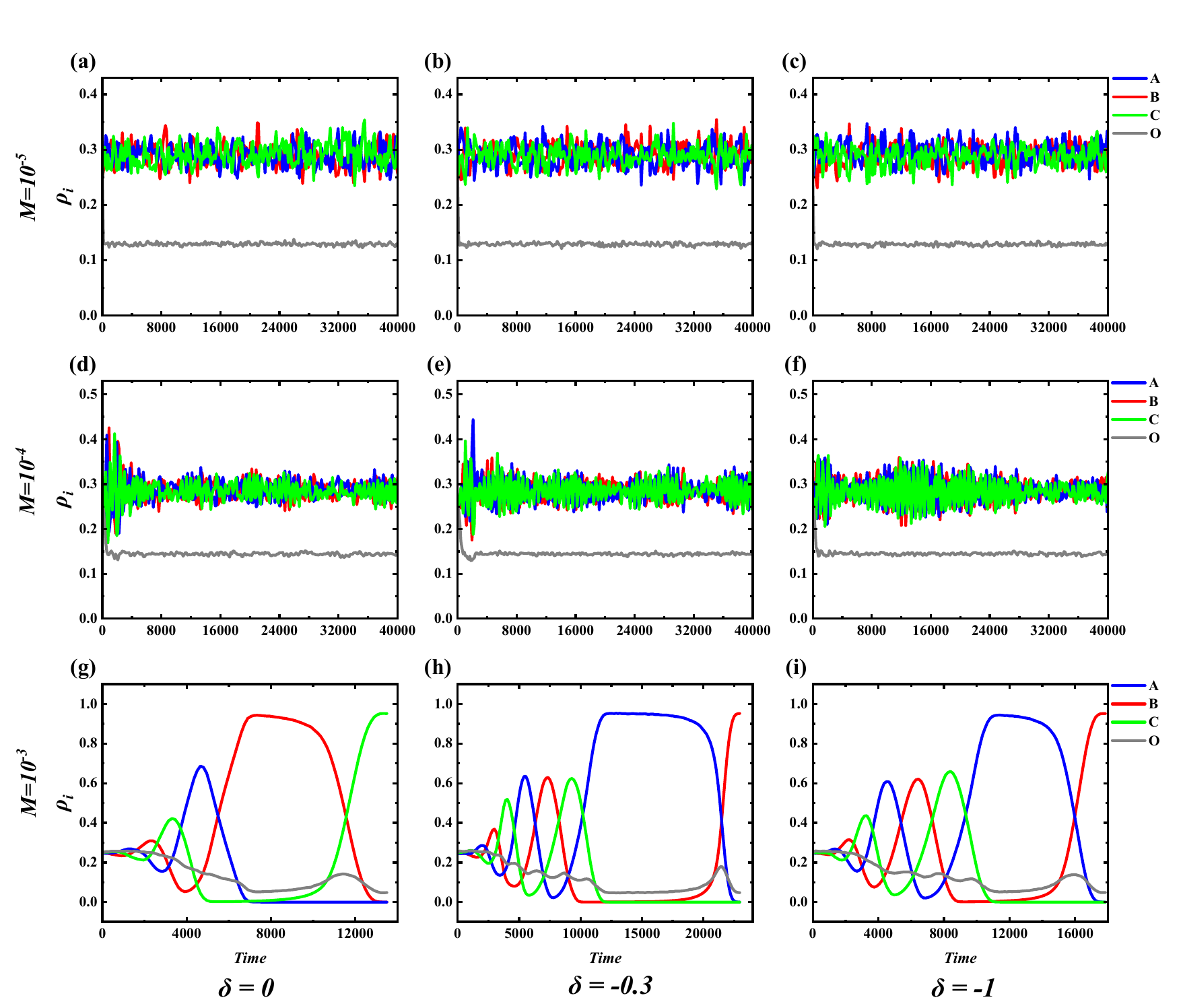}
	\caption{Temporal evolution of species densities under different combinations of mobility $M$ and high-order interaction parameter $\delta$. Panels (a-c) show the time series of the densities for $M=10^{-5}$ and $\delta=0$, $-0.3$, $-1$ respectively; panels (d-f) correspond to $M=10^{-4}$ and $\delta=0$, $-0.3$, $-1$; panels (g-i) correspond to $M=10^{-3}$ and $\delta=0$, $-0.3$, $-1$. In each panel, red, blue, and green lines represent the densities of species $A$, $B$, and $C$, respectively, while the black line indicates the fraction of empty sites. The dynamical trajectories reveal the coexistence or extinction outcomes for each parameter set, consistent with the spatial patterns shown in Fig.~\ref{fig:2}.}
	\label{fig:3}
\end{figure*}

Overall, Figure~\ref{fig:2} shows similar patterns of formations in biodiversity to those in classic RPS systems in the absence of intraspecific competition: As mobility increases, pattern formations and associated biodiversity present similar features: a spirally entangled coexistence that can lead to extinction, where a single species survives.
When $\delta=0$ (i.e., intraspecific competition rate $P=0.1$ and no higher-order interactions), the results reflect those of the conventional RPS scenario~\cite{42}. 
In Figs.~\ref{fig:2}(a-c), under a low mobility rate ($M=10^{-5}$), species $A$, $B$, and $C$ can coexist stably regardless of the intraspecific competition parameter $\delta$, with the spatial patterns exhibiting a fragmented configuration, indicating high species diversity. 
As $\delta$ decreases, subtle variations emerge in the pattern details, but the overall system diversity remains largely unaffected. Ordered spiral waves emerge due to the cyclic competition among species. 
In Figs.~\ref{fig:2}(d-f), under intermediate mobility ($M=10^{-4}$), the spatial structure evolves from a disordered state to an ordered configuration, with distinct spiral wave patterns emerging. Under these conditions, the three species maintain stable coexistence. Visually, the spatial patterns in Figs. 2(d) and 2(f) exhibit a high degree of similarity, suggesting that the macroscopic spiral wave topology is robust to variations in $\delta$ under intermediate mobility. However, in Figs.~\ref{fig:2}(g-i), under high mobility ($M=10^{-3}$), the system exhibits biodiversity loss. When $\delta=0$, species $C$ becomes the sole survivor (Fig.~\ref{fig:2}(g)). As $\delta$ further decreases to $-0.3$ and $-1$, species $A$ dominates the entire space while species $B$ and $C$ go extinct (Figs.~\ref{fig:2}(h,i)). 
These findings demonstrate that weak intraspecific competition accompanied by high mobility significantly undermines the diversity maintenance mechanisms, leading to the dominance of a single species within the system.

To comprehensively analyze the evolutionary dynamics of species, we present the temporal evolution of the densities of three species and empty sites, as shown in typical snapshots in Figs.~\ref{fig:2} and~\ref{fig:3}. Under low mobility ($M=10^{-5}$) (Figs.~\ref{fig:3}(a-c)), the densities of all three species exhibit minor fluctuations within the range of $0.25$ to $0.35$ approximately under the regulation of higher-order interactions, regardless of the value of $\delta$, and they coexist stably without extinction. The proportion of empty sites remains low with minimal fluctuations, indicating near-dynamic stability. A comparison across different $\delta$ shows that when $\delta = 0$, the system's dynamics resemble those of the classic RPS model. 
As $\delta$ decreases, the higher-order effects intensify, yet the overall biodiversity and the proportion of empty sites remain largely unaffected. This suggests that the combined effect of low mobility and strong higher-order interactions enables the system to sustain biodiversity and maintain dynamic equilibrium. Furthermore, the persistent presence of empty sites with stable fluctuations provides the spatial basis for long-term cyclic competition among species.

As shown in Figs. 3(d-f), under intermediate mobility, the species densities exhibit stable oscillations after the system has reached its steady state, with each species fluctuating within a range of approximately $0.25-0.35$. These represent stable limit-cycle oscillations of the system. The density of empty sites shows a slight increase ($0.13-0.17$), and a decreasing $\delta$ leads to stronger higher-order interactions accompanied by a modest increase in the average density of empty sites. In addition, the system exhibits pronounced dynamic periodicity. 
These results demonstrate that enhanced higher-order interactions, mediated by food availability through the regulation of intraspecific competition, effectively suppress the expansion of local dominance. This mechanism creates recovery opportunities for disadvantaged species and vacant spaces, thereby further increasing the complexity of the system's spatial structure.

Under high mobility ($M=10^{-3}$) (Figure~\ref{fig:3}(g-i)), species $C$ eventually dominates the entire space when $\delta=0$, while species $A$ and $B$ become extinct. As $\delta$ decreases to $-0.3$ and $-1$, species $A$ becomes the sole dominant species as its density approaches $1$. Notably, in the high mobility regime, regardless of the value of $\delta$, the long-term evolution of the system exhibits an absolute dominance by a single species. However, the density of vacancies never completely disappears, but instead consistently remains at a statistically significant level, although well below $1$. The results demonstrate that the effect of food availability on intraspecific competition amplifies the competitive advantage of a given species under higher-order interactions. When combined with the spatial constraints inherent in low-mobility conditions, these dynamics ultimately lead the system to single-species dominance.

In summary, at low and intermediate mobility levels, the three species maintain stable coexistence via higher-order regulatory mechanisms, exhibiting spatially complex structures such as fragmented patches or spiral waves. In the temporal evolution of species densities, the densities of all species exhibit stable oscillations, while the proportion of empty sites remains consistently above zero. As mobility increases, higher-order interactions rapidly amplify competitive advantages, leading to monopolization and single-species dominance across the spatial domain. Nevertheless, empty sites persist at minimal yet non-zero levels even under such conditions. This results from the combined effects of spatial and higher-order mechanisms. Collectively, food-mediated higher-order control of intraspecific competition plays a determinant role in shaping both biodiversity persistence and the system's spatiotemporal organization, with blank node dynamics offering fundamental insights into ecosystem homeostasis and resource partitioning.

\begin{figure}[ht]
	\centering
	\includegraphics[width=0.95\linewidth]{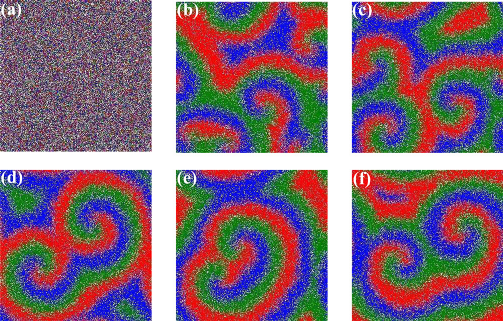}
	\caption{Characteristic snapshots for $\delta=0$ and $ M=10^{-4}$ at various time steps: 
		(a) $1$, (b) $150$, (c) $2,950$, (d) $9,850$, (e) $19,000$, and (f) $180,000$. The color information is the same in Fig.~\ref{fig:2}.
		The initial random distribution (a) gradually develops into characteristic spiral wave patterns (b-f) as the system evolves.}
	\label{fig:4}
\end{figure}

\begin{figure}[ht]
	\centering
	\includegraphics[width=0.95\linewidth]{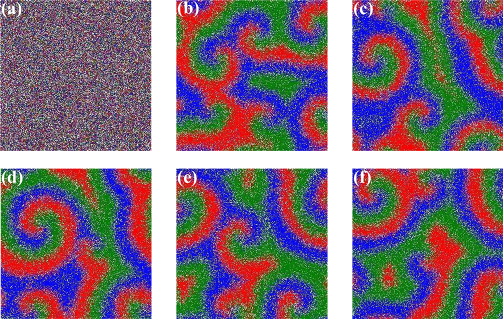}
	\caption{Characteristic snapshots for $\delta=-0.3$ and $M=10^{-4}$ at various time steps: (a) 1, (b) 150, (c) 2,950, (d) 9,850, (e) 19,000, and (f) 180,000. The color information
		is the same in Fig.~\ref{fig:2}.The initial random distribution (a) gradually develops into characteristic spiral wave patterns (b-f) as the system evolves.}
	\label{fig:5}
\end{figure}

\begin{figure}[ht]
	\centering
	\includegraphics[width=0.95\linewidth]{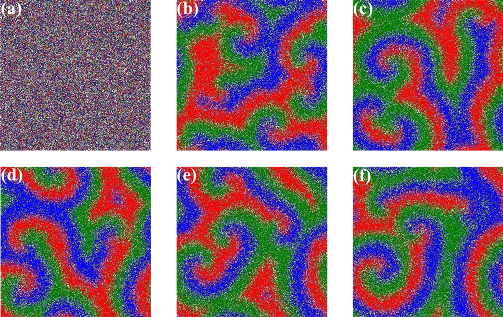}
	\caption{Characteristic snapshots for $\delta=-1$ and $M=10^{-4}$ at various time steps: (a) 1, (b) 150, (c) 2,950, (d) 9,850, (e) 19,000, and (f) 180,000. The color information is the same in Fig.~\ref{fig:2}.
		The initial random distribution (a) gradually develops into characteristic spiral wave patterns (b-f) as the system evolves.}
	\label{fig:6}
\end{figure}

From Figs.~\ref{fig:2} and~\ref{fig:3}, we discuss the evolution of species across various values of $\delta$. To observe microscopic insights into the evolutionary dynamics across three distinct $\delta$ conditions, we analyze characteristic snapshots at different time points as illustrated in Fig.~\ref{fig:4}. The initial distributions for various conditions, each species, and empty space, accounting for about $25\%$ of the total, are depicted. When $\delta=0$ (see Fig.~\ref{fig:4}), where intraspecific competition remains unregulated by food availability and higher-order interactions show negligible effects, the system exhibits characteristic spiral spatial patterns with all three species maintaining stable long-term coexistence, reflecting classical RPS cyclic dynamics. In Figs.~\ref{fig:5} and~\ref{fig:6}, with $\delta$ reduced to $-0.3$ and $-1$, enhanced higher-order interactions result in narrower spiral waves and a systematic decrease in the spatial scale of the patterns (for quantitative analysis, see Sec. 3.2). The results demonstrate that food-mediated regulation of intraspecific competition further amplifies the accumulation of local dominance and differentiation in scale.
However, under intermediate mobility ($M=10^{-4}$), the three species maintain dynamic coexistence over extended timescales, regardless of the $\delta$ values, with no observed extinction or monodominance. Notably, as the higher-order regulatory parameter $\delta$ decreases, the average density of empty sites increases slightly, primarily in boundary regions. The persistence of empty sites provides essential spatial opportunities for cyclic competition and ecological recovery.

\begin{figure}[ht]
	\centering
	\includegraphics[width=0.95\linewidth]{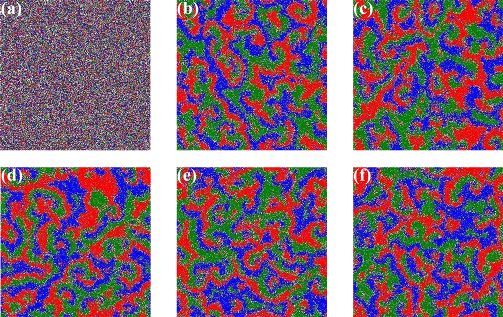}
	\caption{Characteristic snapshots for $\delta=-1$ and $M=10^{-5}$ at various time steps: (a) 1, (b) 150, (c) 2,950, (d) 9,850, (e) 19,000, and (f) 180,000. The color information is the same in Fig.~\ref{fig:2}.
		The initially random distribution (a) gradually evolves into stable, fine-grained coexistence patterns, accompanied by significant spatiotemporal fluctuations, as the system develops (b-f).}
	\label{fig:7}
\end{figure}

Under low mobility ($M=10^{-5}$), with $\delta=-1$, the RPS system maintains its characteristic diversity coexistence pattern. The three species remain well-mixed globally, while the spatial structure preserves a fine-grained patchy configuration. Regardless of $\delta$, the system sustains empty sites at consistently low densities with uniform species mixing, showing no mono-dominance. High mobility enhances species dispersal and mixing, effectively suppressing the amplification of local competitive advantages.

\begin{figure}[ht]
	\centering
	\includegraphics[width=0.95\linewidth]{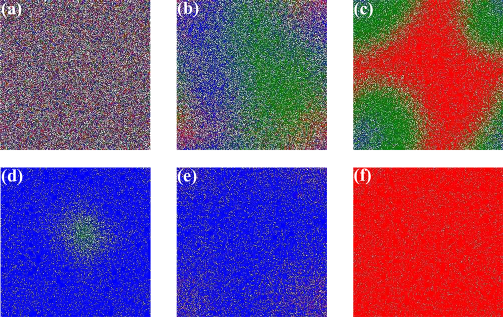}
	\caption{Characteristic snapshots for $\delta=-1$ and $M=10^{-3}$ at various time steps: 
		(a) 1, (b) 150, (c) 2,950, (d) 9,850, (e) 19,000, and (f) 180,000. The color information is the same in Fig.~\ref{fig:2}. 
		Starting from a random initial distribution (a), the system exhibits rapid loss of biodiversity, evolving into single-species dominance states (d-f) as time progresses.}
	\label{fig:8}
\end{figure}

Under high mobility conditions ($M=10^{-3}$), with $\delta=-1$, the spatial structure of the RPS system undergoes a dramatic transition from an initial mixed state to large-scale domain formation, culminating in complete mono-dominance by a single species. During this evolutionary progression, one species rapidly expands and eventually occupies the entire spatial domain. In contrast, other species and empty sites progressively diminish, ultimately resulting in extensive monochromatic patches dominated by the winning species. This final state is characterized by complete loss of biodiversity, disappearance of complex spatial patterns, and significantly reduced ecological resilience and recovery capacity.

In summary, reduced mobility significantly compromises the spatial mixing capacity of the three species and their ability to maintain diversity. Under high mobility conditions, the system consistently exhibits stable biodiversity and cyclic patch dynamics regardless of variations in higher-order regulation. In contrast, the combined effects of low mobility and strong higher-order interactions dramatically amplify regional competitive advantages, driving the system into a terminal state characterized by mono-species dominance and pattern polarization.
Furthermore, varying values of $\delta$ have a significant impact on the proportion of empty sites.

As shown in Fig.~\ref{fig:2}, it is clear that the strength $\delta$ significantly affects the dilution of the spiral wave. 
To further elucidate these phenomena quantitatively and advance our central objective of understanding how higher-order interactions modulate biodiversity through intraspecific competition, we evaluate basin entropy $S_b$, a measure that quantifies the unpredictability of the final state in dynamical systems~\cite{43,44}.
To compute basin entropy, we initially divide the network into $L_b \times L_b$ non-overlapping sub-networks and subsequently calculate the Gibbs entropy of each sub-network using the following equation:
\begin{equation}
S_i = -\sum_{j=1}^{N_A} p_{ij} \log p_{ij}.
\end{equation}
Here, $N_A$ denotes the number of four states (three species and empty sites) in this model, i.e., $N_A = 4$, and the term $p_{ij}$ denotes the proportion of sites occupied by species $j$ within the $i$-th box. Specifically, $p_{ij} = 0$ if the species inside the box is species $j$. 
The values of basin entropy lie within the interval $[0,\log(N_A)]$, where $N_A$ denotes the total number of distinct possibilities within the lattice. A basin entropy of zero corresponds to a scenario in which a single attractor characterizes the system. Conversely, a basin entropy of $\log(N_A)$ indicates a scenario where the basins are fully randomized, with $N_A$ attractors that are equally probable. 
Then, basin entropy $S_b$ is calculated by
\begin{equation}
S_b = \dfrac{S}{N_b},
\end{equation}
where $S$ is given by the sum of all $S_i$:
\begin{equation}
S = \sum_{i=1}^{N_b} S_i,
\end{equation}
with $N_b$ of the number of boxes of size $L_b \times L_b$,
and the result of $S_b$ across a range of $\delta$ is presented in Fig.~\ref{fig:9}(a).

\begin{figure}[ht]
	\centering
	\includegraphics[width=1\linewidth]{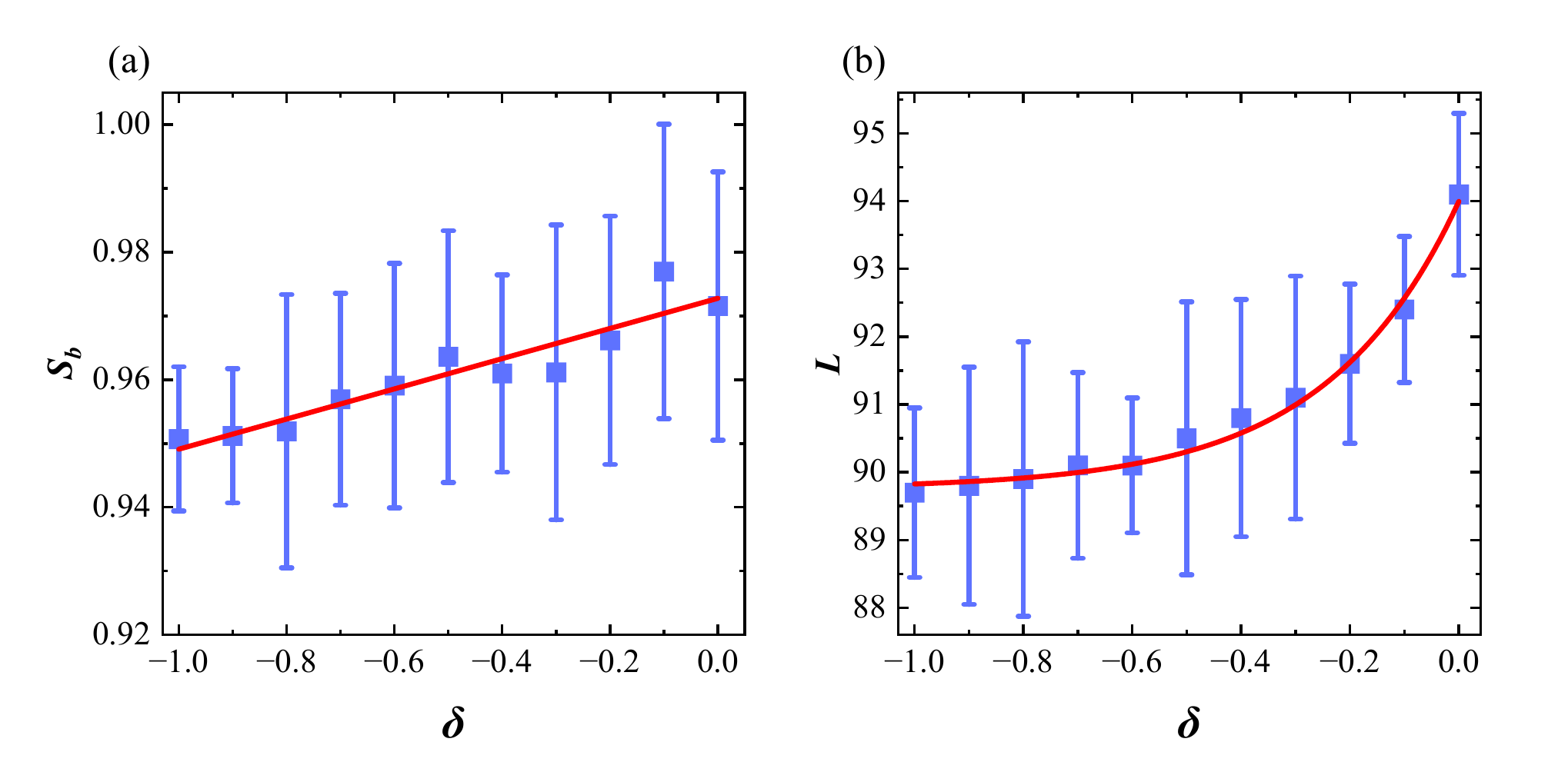}
	\caption{\label{fig:9}(a) Basin entropy $S_b$ as a function of $\delta$ at $M=10^{-4}$ with different values of $L_b=5$. The box denotes basin entropy, while the error bars indicate the standard deviation derived from $300$ independent simulations.
		(b) Characteristic wavelength $L$ as a function of $\delta$ at $M=10^{-4}$. The data points represent averages with error bars indicating the standard deviation. The red solid line represents the exponential fit ($R^{2} = 0.993$), showing that the spatial wavelength decreases exponentially and saturates as higher-order interactions strengthen.}
	\label{fig:9}
\end{figure}

\begin{figure*}[ht]
	\centering
	\includegraphics[width=0.7\linewidth]{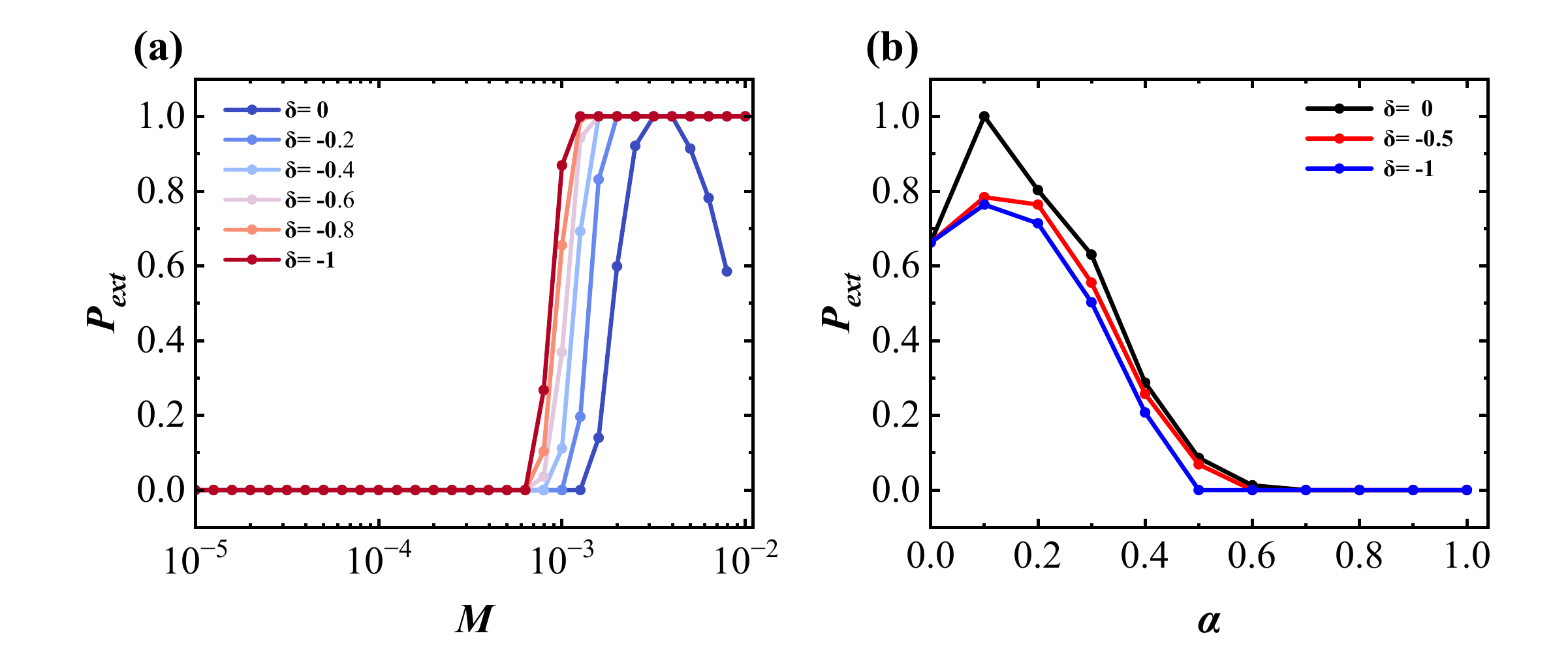}
	\caption{\label{fig:10}(a) The extinction probability $P_{ext}$ varies with $M$ across different values of $\delta$. 
		(b) The extinction probability $P_{ext}$ varies with $\alpha$ across different values of $\delta$ for fixed mobility $M=10^{-4}$. 
		The network size is $300\times300$.}
\end{figure*}

The quantitative analysis in  Fig.~\ref{fig:9}(a) reveals a statistically robust decrease in basin entropy $S_b$ as the higher-order interaction strength increases (i.e., as $\delta$ becomes more negative). This trend indicates that stronger HOI suppress spatial disorder and enhance the predictability of the system's configuration.

Within this parameter regime, based on $300$ independent realizations, the entropy $S_b$ exhibits a robust linear dependence on $\delta$. A linear regression analysis yields the following relation:
\begin{equation}
S_b = 0.973 + 0.024\delta,
\label{eq:linear_reg}
\end{equation}
where the intercept is $\alpha_0 = 0.973$ (95\% CI: $[0.969, 0.977]$) and the slope is $\beta = 0.024$ (95\% CI: $[0.0169, 0.0304]$). This positive slope confirms that stronger higher-order interactions (more negative $\delta$) lead to a statistically significant reduction in basin entropy.

To further test the characteristics of the pattern, we simultaneously tested the relevant wavelengths; the pertinent details are shown in Fig.~\ref{fig:2}(e). Mathematically, $L$ is defined as the average spatial period of the spiral waves. We derived this value by measuring the linear distance between consecutive wave fronts of the same species along randomly selected cross-sections of the lattice. The reported $L$ values in Fig.~9(b) are averages obtained from multiple measurements across independent realizations to minimize statistical error. Fig.~9(b) shows a concurrent decrease in the characteristic wavelength $L$ under stronger HOI. A nonlinear regression analysis reveals that this dependence follows an exponential decay model ($R^2 = 0.993$). The relationship is described by $L(\delta) = L_{0} + A \exp(\delta / \tau)$, where the baseline wavelength saturates at $L_{0} \approx 89.8$ (95\% CI: $[89.5, 90.0]$) and the characteristic decay scale is $\tau \approx 0.24$ (95\% CI: $[0.20, 0.29]$). This nonlinear relationship demonstrates that the enhancement of spatial order is physically manifested as a structural refinement of spiral patterns into more compact domains, which rapidly stabilizes as higher-order interactions strengthen. Notably, although $S_b$ values remain within a constrained interval ($0.94$--$0.97$), this systematic reduction in both entropy and wavelength demonstrates that HOI actively reinforce spatial order beyond the baseline stability provided by species mobility and cyclic dominance. The emergence of these finely-structured yet highly organized patterns represents a key mechanism through which HOI contribute to the long-term stability and coexistence of species.

\subsection{\label{sec:macro_sim}Macroscopic phenomena in the evolutionary dynamics of species}

In classic systems of cyclic competition on spatially extended systems, it has been found that the biodiversity of species, whether through coexistence or extinction, can be sensitively influenced by an individual’s mobility. Our previous study systematically analyzed the effects of different $\delta$ values on spatial structures under the intermediate mobility condition ($M=10^{-4}$), As shown in Fig.~\ref{fig:9}, both the proportion of vacant sites and spatial entropy remain essentially constant across the entire range of $\delta$ values, indicating that the regulatory effect of $\delta$ on intraspecific competition intensity has limited influence on the system's macroscopic spatial structure within this mobility regime. Notably, the system maintains its characteristic self-organized pattern structure and stable empty site levels, regardless of variations in $\delta$, suggesting a nuanced role of higher-order interactions in this regard. Thus, we may argue that species may have a greater chance of coexisting through HOI, even if such a finding is obtained when considering medium mobility. Regarding typical snapshots, since the effect of HOI on biodiversity across a broad spectrum of mobility regimes is ambiguous, we investigate this effect using the extinction probability. To obtain the robust and reasonable result, we consider the $800$ random and independent realizations and the associated result for different values of $\delta$, illustrated in Fig.~\ref{fig:10}.

Figure~\ref{fig:10}(a) demonstrates the variation in extinction rate ($P_{ext}$) with respect to mobility rate ($M$) under different $\delta$ values. When $M < 10^{-4}$, the extinction probability remains consistently near zero regardless of $\delta$ values, indicating a highly stable system state. As the mobility rate $M$ increases, the extinction probability undergoes a rapid transition to unity at relatively low $M$ values, causing the system to enter a high-risk extinction regime at an earlier stage. In contrast, systems with larger $\delta$ (i.e., weaker HOI) values demonstrate an enhanced capacity to maintain biodiversity and stability even at elevated mobility rates.

Figure~\ref{fig:10}(b) illustrates the relationship between the extinction probability $P_{ext}$ and the parameter $\alpha$ under different $\delta$ values. It is observed that for small values of $\alpha$, regardless of $\delta$, the system exhibits a high extinction probability, indicating that species diversity is in a highly fragile state. As $\alpha$ increases, $P_{ext}$ decreases sharply, and for $\alpha>0.5$, all curves approach zero, suggesting that strong higher-order interactions promote long-term species coexistence.

\section{\label{sec:conc}Conclusion}

Preserving biodiversity is a crucial task in ecosystem conservation. However, as species interactions become more complex, traditional models and methodologies often fail to capture essential features of ecosystems. To address this challenge, this study extends the classic Rock-Paper-Scissors model by incorporating higher-order interactions to explore their effects on species coexistence, community stability, and extinction probabilities~\cite{38,31}.

In our model, the competitive interactions between species are significantly influenced by higher-order interactions, particularly through variations in the strength of intraspecific competition. 
Crucially, our quantitative analysis reveals that stronger HOI induce a 'structural tightening' of the spatial patterns. Rather than simply transitioning from disorder to order, the system organizes into more compact, high-efficiency domains (shorter wavelengths), which enhances structural robustness.
The simulation results show that under low and intermediate mobility conditions, the system maintains stable species coexistence, characterized by the emergence of complex spatial structures such as spiral waves. In these regimes, biodiversity is robust to variations in interaction strength. In contrast, under high mobility conditions, the spatial patterns are destabilized, leading to a significant decline in biodiversity and the emergence of single-species dominance. This highlights that excessive mobility, rather than low mobility, jeopardizes ecological balance by homogenizing the population distribution.
Additionally, we further analyzed the impact of HOI on extinction probabilities. The results indicate that the extinction probability of species is significantly altered under the influence of higher-order interactions, especially under strong higher-order interactions (large negative $\delta$), which shift the extinction threshold toward lower mobility values, thereby increasing the extinction risk. This finding contrasts with the predictions of traditional RPS models and underscores the critical role of HOI in shaping ecological dynamics~\cite{46}.

In conclusion, this study demonstrates that higher-order interactions play a pivotal role in maintaining species coexistence and community stability. By regulating intraspecific competition and resource distribution, HOI significantly alter the competitive relationships between species and the dynamic features of ecosystems. As our understanding of ecosystem complexity continues to evolve, future research can further investigate how HOI affect species diversity and stability across various ecosystems, providing new theoretical insights for biodiversity conservation and species management.

\section*{Acknowledgements}

C. P. is funded by the Yunnan Fundamental Research Projects (No.202401AU070018), the Scientific Research Fund of the Yunnan Provincial Department of Education (No. 2024J0774). 
Y.K. acknowledges support from the Yunnan Fundamental Research Projects (Grant Nos. 202501AU070193), the Scientific Research Fund of the Yunnan Provincial Department of Education (Grant No. 2025J0579), Yunnan University of Finance and Economics Fund(2025D60) and the Foundation of Yunnan Key Laboratory of Service Computing (Grant No. YNSC24125).
J.P. was supported by the National Research Foundation of Korea (NRF) grant funded by the Korea government (MSIT)(No. RS-2023-NR076590). J.P. was also supported by Global-Learning \& Academic research institution for Master's $\cdot$ PhD students, and Postdocs(G-LAMP) Program of the National Research Foundation of Korea(NRF) grant funded by the Ministry of Education(No. RS-2025-25442355).

\bibliographystyle{elsarticle-num}
\bibliography{references}

\newpage
\appendix 

\section{Supplementary Material A: ODD Protocol} 
	
	\subsection*{1. Purpose and Patterns}
	The primary purpose of this model is to investigate how higher-order interactions, implemented through a resource-mediated intraspecific competition mechanism, affect species coexistence, spatial pattern formation, and extinction probability in a spatial Rock-Paper-Scissors system. The model is designed to reproduce and explain key ecological patterns, including the formation and stability of spiral waves under low and intermediate mobility regimes, the loss of biodiversity under high mobility conditions, and the stabilizing role of HOIs in delaying extinction thresholds and modulating spatial characteristic scales.
	
	\subsection*{2. Entities, State Variables, and Scales}
	The model comprises agents operating on a two-dimensional square lattice with periodic boundary conditions. Grid cells represent the basic spatial units, where each cell can be occupied by species $A$, $B$, $C$, or remain empty ($\emptyset$). The system dynamics are governed by state variables including the mobility rate ($M$), baseline intraspecific competition rate ($\alpha$), and the higher-order modulation factor ($\delta$). Crucially, the local food resource abundance ($\rho$) is implemented as a dynamic spatiotemporal variable, defined instantaneously as the count of empty sites within the six nearest-neighbor positions surrounding any interacting pair. This variable updates continuously as the lattice configuration evolves through ecological processes.
	
	\subsection*{3. Process Overview and Scheduling}
	Model processes follow an asynchronous random sequential updating scheme. In each elementary simulation step, a focal individual and one of its random neighbors are selected, upon which one of four fundamental ecological processes is probabilistically chosen based on their states. The unit of time is the Monte Carlo Step (MCS), where 1 MCS corresponds to $N$ elementary steps. All simulations run for a duration of $T = 32N$ MCS to ensure the system bypasses transient dynamics and reaches a quasi-steady state before any data collection occurs.
	
	\subsection*{4. Design Concepts}
	Global patterns such as spiral waves and biodiversity transitions emerge spontaneously from local stochastic interactions between individuals. Agents effectively sense their local environment through the computation of $\rho$ values within their immediate interaction neighborhood. All interactions occur through localized stochastic events following defined transition probabilities, with the selection of individuals, neighbors, and actions all being random processes. Data collection focuses on system-level observations including species densities, basin entropy, and spatial correlation lengths measured after the initial equilibration period.
	
	\subsection*{5. Initialization}
	The lattice is initialized with a random uniform distribution where each cell is independently assigned to species $A$, $B$, $C$, or the empty state ($\emptyset$) with equal probability (25\% each). All model parameters ($\sigma$, $\mu$, $M$, $\alpha$, $\delta$) remain fixed throughout each simulation run as described in the main text, ensuring consistent initial conditions across different parameter sets.
	
	\subsection*{6. Input Data}
	The model operates as a self-contained theoretical framework that does not incorporate external time-series input data to represent time-varying processes. All system dynamics emerge exclusively from the internal interaction rules and initial configuration, maintaining theoretical consistency and allowing for clear mechanistic interpretation of the results.
	
	\subsection*{7. Submodels}
	The model implements four core ecological processes through precisely defined submodels. Interspecific competition follows classic rock-paper-scissors dynamics ($XY \to X\emptyset$) at rate $\sigma$. Reproduction occurs when individuals colonize adjacent empty sites ($X\emptyset \to XX$) at rate $\mu$. Mobility involves spatial mixing through position swapping between neighbors ($X\square \to \square X$) at rate $r = 2MN$. The core innovation lies in the intraspecific competition mechanism ($XX \to X\emptyset$), where the rate $P$ is dynamically computed as:
	\begin{equation}
	P = \alpha \exp\left(\frac{\rho}{6}\delta\right),
	\end{equation}
	with $\rho \in [0, 6]$ representing the instantaneous count of empty sites in the specific six-cell neighborhood of the interacting pair. This formulation ensures that competition intensity responds non-linearly to local resource availability, implementing the key higher-order interaction mechanism.

\end{document}